# A Human–LLM Teaming Framework for Privacy Risk Analysis: An Illustration with CBDC-Based Welfare Schemes

**Sourya Joyee De (Indian Institute of Technology Kanpur, India; Email: sourya@iitk.ac.in)**
**Abdessamad Imine (Université de Lorraine, , CNRS, Inria, LORIA F-54000 Nancy, France; Email: abdessamad.imine@loria.fr)**
**Abstract:** Central Bank Digital Currency (CBDC)-based welfare schemes may be potentially privacy invasive as they process significant volumes of beneficiary personal data and lead to privacy harms such as surveillance, discrimination and stigmatization. Such welfare delivery schemes involve complex digital ecosystems and large number of stakeholders. Consequently, to examine their privacy risks, privacy risk assessments require extensive information gathering and synthesis, complex reasoning, scenario explorations, contextual evaluation and human judgement. Thus, they present ideal scenarios for human-LLM teaming, where effective integration of complementary human and LLM capabilities can yield an outcome far superior to either human-only or LLM-only assessments. In this paper, we propose a first human–LLM teaming framework for the systematic privacy risk analysis methodology called PRIAM. The framework specifies an iterative collaborative process in which the LLM processes large-scale documentary evidence to produce initial outputs, which are then interpreted and evaluated by human experts who direct their further refinement by the LLM and exercise their judgement to finalize the output. We illustrate the framework on the data characterization activity of PRIAM using a CBDC-based welfare scheme use case. The illustration demonstrates that while LLMs generate the initial data categories and assign initial values to data attributes, human experts evaluate and provide feedback to refine them, distinguishing documented evidence from inferences, identifying information gaps, and flagging unsupported or ambiguous outputs. This framework serves as a foundational contribution towards human-AI teaming for privacy risk assessments.

## 1. Introduction

Central Bank Digital Currency (CBDC), a digital counterpart of government-backed fiat money, has opened a new avenue for delivering efficient, targeted and accountable digital welfare. The Reserve Bank of India has conducted multiple pilots for welfare schemes linked to India's CBDC, eRupee, across its states and union territories, such as Gujarat, Puducherry, and Chandigarh, where beneficiaries receive food subsidies in digital rupee.[1] Features of eRupee, such as traceability, programmability, and the availability of detailed transaction records, enhance the targeting and accountability of these welfare schemes, yet pose potentially severe privacy risks, including financial and behavioural profiling, surveillance, stigmatization, and discrimination against beneficiaries. Programmability, for example, allows the sponsoring entity to ensure that beneficiaries use allocated funds for a specific purpose, preventing misuse[2], but may also enable reconstruction of transaction history and inferences about beneficiaries' economic circumstances. Comprehensively identifying these privacy risks requires a privacy risk assessment that systematically examines the data processed, the digital infrastructure, and the stakeholders involved to identify potential weaknesses in implemented data protection mechanisms and entities that could exploit them, resulting in unacceptable privacy incidents and harms. However, applying privacy risk analysis (PRA) methodologies in a complex digital ecosystem, such as CBDC-based welfare schemes, necessitates substantial information gathering, synthesis, risk exploration and scenario analysis (Mollaeefar et al., 2025; Bissoli et al., 2026). PRAs are resource-intensive, error-prone, and heavily dependent on expert knowledge and experience (Mollaeefar et al., 2025; Bissoli et al., 2026). Large Language Models (LLMs) offer capabilities to augment various activities within PRAs through large-scale data gathering and synthesis, relational reasoning, scenario generation and exploration of risks and mitigation strategies. However, the outcomes of PRA exercises have significant human rights and legal implications. Thus, PRA methodologies such as PRIAM (De & Le Métayer, 2016) continue to require human oversight and present an ideal case for human-LLM teaming, in which human intuition and contextual knowledge complement the computational capabilities and data processing power of AI.

---

[1] RBI Expands Digital Rupee Use for Welfare and Cross-Border Payments, ETBFSI
[2] DigitalRupee09012025.pdf

Drawing on Complementary Team Performance (CTP) theory (Hemmer et al., 2025), we envision that the performance of PRAs can significantly improve when LLM capabilities and human expertise are well coordinated to meet the demands of individual PRA activities. Privacy engineering investigation into the systematic integration of AI and the potential of human-AI teaming for privacy threat modelling and impact assessments is in a nascent stage (Mollaeefar et al., 2025; Bissoli et al., 2026; Lee et al., 2026; Lee, 2026). At present, there is no comprehensive framework to guide the collaboration between human experts and AI agents across PRA activities.

Our contributions are twofold. First, we develop a framework that specifies how human experts and LLMs can work together across PRIAM activities, with the LLM generating initial analyses from available evidence and the human expert interpreting, evaluating, directing refinement, and exercising final judgment. Second, we demonstrate how a PRIAM activity (such as data characterization) can be performed through this teaming framework in the context of CBDC-based welfare delivery, illustrating the iterative interaction between the LLM and human expert. Through this illustration, we show how LLM capabilities can augment, rather than replace, human expertise in PRA activities while maintaining human oversight.

The rest of this paper is organised as follows: Section 2 describes a case study for CBDC-based Welfare Scheme and its potential privacy risks; Section 3 proposes a human-LLM teaming framework for the privacy risk analysis methodology PRIAM; Section 4 illustrates data characterization, a PRIAM activity, using the human-LLM teaming framework for the CBDC-based Welfare Scheme case study; Section 5 discusses related works; finally, Section 5 concludes with future directions.

## 2. Privacy Risks of CBDC-based Welfare Schemes

As a case study, we consider a welfare program targeting women, in which an eligible beneficiary receives their annual payout in eRupee in their digital wallet and can spend it via

QR codes, transfer it to their bank account, or withdraw cash[3]. This welfare scheme would involve various stakeholders, including the scheme administrator, government agencies responsible for eligibility verification, banks and other payment intermediaries, the unique identity infrastructure, digital service and application providers, and enrolment and verification intermediaries. The digital infrastructure includes an application for identity verification and scheme management; the corresponding backend database; a unique identity infrastructure for identity verification; government databases and verification systems to validate eligibility and beneficiary information; banking, direct benefit transfer, and CBDC infrastructure for the direct transfer of eRupee to beneficiaries. When a beneficiary decides to spend the eRupee, the transaction, its value, time, location, recipient, merchant, as well as the purpose of the transaction, may be recorded. These transactions can be linked to beneficiary identities; multiple transactions may be combined to create longitudinal profiles, and inferences can be drawn from these profiles. Without appropriate privacy-preserving measures in place, these profiles may be accessed by multiple stakeholders involved in the scheme, even if they do not need to process this data, as well as by unauthorized parties who are not stakeholders in the scheme. Ultimately, in the absence of appropriate data protection measures, these profiles may potentially be used for purposes beyond the original welfare objective.

## 3. A Framework for Human-LLM Teaming for PRIAM

### 3.1 Human-LLM Complementarity in PRIAM

Despite PRIAM (De & Le Métayer, 2016) being a structured methodology, its application requires human experts to perform a range of cognitively demanding activities. The first phase of PRIAM involves an exhaustive process of gathering information about the data processing activity in question from heterogeneous sources and identifying relevant entities, categories and attributes. This phase, therefore, presents the challenge of not overlooking any information that could significantly affect the risk assessment. The second phase, which is the risk assessment phase, requires the generation of candidate risk scenarios in the form of harm trees, reasoning about relationships between privacy weaknesses and risk sources on one hand and feared events and harms on the other, assigning appropriate input likelihood values for a risk source exploiting a privacy weakness and assigning appropriate severity values depending on the impact level on a potential target group. These harm scenarios, along with severity and likelihood assignments, must be evidence-based; without them, the reliability of the privacy risk analysis process will be questionable. Thus, both phases require considerable information processing and contextual judgment. LLMs can potentially contribute to activities that necessitate broad information processing, synthesis, candidate scenario generation, and exploration of alternative scenarios. On the other hand, human experts can contribute to the evaluation of LLM outputs, providing feedback to refine outputs, judging the final output – all based on their domain knowledge, contextual interpretation, and comprehension of stakeholder perspectives - and ultimately remaining accountable for the results of the risk assessment. Thus, human experts and LLMs can act as complementary team members, with the output of one reviewed by the other for possible improvements, and the teamwork ultimately produces an outcome far superior to what either team member could have achieved on their own. An LLM may generate a broad set of candidate privacy risks based on documents on CBDC and CBDC-based welfare schemes from heterogeneous sources. Human analysts can eliminate implausible risks, add risks that the LLM has overlooked, identify context-specific omissions, challenge assumptions and evidence used by the LLM and provide additional assumptions. The refined information can then be supplied to the LLM to explore additional risk pathways. Thus, LLM's ability to synthesize large volumes of information and human contextual knowledge can operate as mutually reinforcing resources. In the remainder of this section, we discuss how LLMs and human experts can serve as team members across different PRIAM activities.

---

[3] Subhadra first govt scheme to offer digital rupee payment - The Times of India

### 3.2 Information Gathering

The first phase of PRIAM involves information gathering, which includes specifying the attributes and categories of seven components – system, stakeholders, data, risk sources, privacy weaknesses, feared events, and privacy harms. A corpus of documents comprising scheme guidelines and official documentation, scheme FAQs, government circulars, RBI CBDC-related documents, application privacy policy, terms and conditions, payment-related documentation, etc., can serve as the knowledge base for an LLM. Based on this, the LLM can perform different actions to characterize the system, stakeholders and data being processed as well as identify potential privacy weaknesses, risk sources, feared events, and harms. For example, it can extract functional specifications of the system, identify actors and supporting assets, and reconstruct data flows to characterize the system. The resulting system representation must then be validated by human experts to ensure its contextual accuracy and completeness. The human role, therefore, is to review the system description and check whether all parts of the system have been accurately mapped and whether enough documented evidence exists to create this system description. Each activity may go through several iterations, with LLMs continually refining their outputs based on human input. For example, the LLM may generate a preliminary system specification. Based on this preliminary specification, the human expert may provide input to the LLM on where the specification lacks completeness and/or which parts may need to be revised due to inaccuracies and provide accompanying documented evidence. The LLM may then generate a revised system specification based on these inputs. Ultimately, the human expert's judgement leads to the final system specification. This final system specification is then used in subsequent activities that depend on it, for example, privacy weakness identification.

To generalize, for each PRIAM activity, the LLM first produces a preliminary output by analyzing, identifying, synthesizing, and reasoning over the documentary evidence available to it. The human expert then interprets and evaluates the output for accuracy, consistency, completeness, and sufficiency of evidence. Based on this evaluation, the human directs the LLM to correct, extend, clarify, or further investigate the output where necessary. The LLM subsequently refines the output in response to this direction, after which the human expert conducts a further evaluation. This process may be repeated iteratively until the human expert judges that the output is sufficiently accurate, complete, consistent, and evidence-supported to be finalized and incorporated into later activities of the privacy risk analysis. Table 1 presents the first set of LLM outputs for each PRIAM activity based on documented evidence. These outputs are then interpreted and evaluated by the human expert, who then provides directions to the LLM for refinement.

**Table 1. Initial LLM Contributions for PRIAM Activities within the Information Gathering Phase**

| PRIAM activity | Initial LLM Contributions based on Documented Evidence |
|---|---|
| System characterization | Extract functional specifications; identify actors; reconstruct data flows; summarize interfaces; identify supporting assets; generate system description (e.g., Data Flow Diagram) |
| Stakeholder characterization | Identify actors across documents; classify potential stakeholder roles; identify relationships |
| Data characterization | Identify data categories; extract attributes such as purpose, retention, visibility and sensitivity |
| Risk source identification | Generate candidate risk sources; identify capabilities (e.g., access rights) and potential motivations from documentation |
| Privacy weakness identification | Identify candidate weaknesses from the system specification and data attributes. |
| Feared event identification | Generate candidate feared events from privacy weaknesses and data flows |
| Privacy Harm identification | Generate possible consequences and affected groups |

### 3.3 Risk Assessment Phase

The risk assessment phase utilizes the outputs of the activities under the information gathering phase to perform two further activities that include harm tree generation and risk level assessment. Harm tree generation requires combining several risk sources, privacy weaknesses and feared events to assess the likelihood of each privacy harm documented during the information gathering phase. Thus, an LLM engages in relational and combinatorial exploration to generate multiple harm trees, whereas the human expert evaluates the plausibility of each branch within each tree, seeks evidence supporting each branch of the tree, investigates whether a particular risk source is capable of exploiting a particular weakness in practice and looks for missing, redundant or alternative nodes. Based on this evaluation, the human expert provides directions for refinement to the LLM and, based on the subsequent LLM output, judges whether the output is sufficiently accurate, complete, consistent, and evidence-supported to be finalized and used for risk level assessment. The LLM generates an evidence-grounded preliminary risk assessment and applies the prescribed risk-computation rules, while the human evaluates the evidence, assumptions, scoring, and resulting risk level and exercises final judgment. Based on the final harm tree for a given harm, the LLM assigns evidence-based likelihood values of risk sources exploiting a privacy weakness. The human experts evaluate these assignments and requests for LLM-based refinements. Finally, the LLM computes the likelihood of the harm based on the final likelihood assignments. The LLM also assigns preliminary severity values for each harm based on evidence about the target group it affects and the extent of the impact. The human expert may then request an LLM-based refinement of the severity values to arrive at the final severity value for each harm. The final judgement about the assigned risk (likelihood and severity) lies with the human expert.

## 4. Illustration using CBDC-based Welfare Schemes

In this section, we illustrate the human-LLM PRIAM activity data characterization using the hypothetical case study of CBDC-based Welfare Scheme discussed in Section 2. Based on available documentation of this women's welfare scheme from various sources, such as government websites, an LLM may identify the categories of data being processed and their attributes, such as sensitivity, volume, origin, purpose, and retention, as specified in PRIAM. The first set of LLM outputs for data categories and attributes is shown in Tables 2 and 3. Evaluating these outputs, the human expert might request that the LLM provide the basis for classifying transaction data as high-sensitivity and ask it to distinguish individual transaction data from transaction history. Based on this input, the LLM provides the justification that a transaction record combining transaction value, location, recipient, merchant, and purpose can reveal information about an individual's economic circumstances and behaviour; hence, transaction records are sensitive in nature. It may also add transaction history that combines several transaction records for a beneficiary as a separate data category. It will also show that the duration of retention cannot be established from available evidence. So, the LLM extracts data categories, assigns preliminary values to data attributes based on the evidence identified, and distinguishes documented facts from its inferences and unknowns. On the other hand, the human expert interprets the LLM's preliminary output, checks for accuracy and completeness, challenges unsupported attribute values, identifies missing information, and directs refinement. Consequently, the LLM reassesses the attribute values, retrieves, organizes and presents supporting evidence, revises characterization and flags unresolved gaps. Ultimately, the human expert finalizes the outputs of the data characterization activity.

**Table 2. Initial LLM Output for Data Categorization**

| Data category | Examples | Source/basis |
|---|---|---|
| **Beneficiary identity data** | Name, unique identity/identifier | Identity verification and beneficiary identification are needed for the welfare scheme to function |
| **Eligibility data** | Eligibility status, scheme-related information | Government databases and verification systems operate together to verify beneficiary eligibility for the scheme |
| **Payment data** | Bank account details | Benefits are transferred to the bank account for transfer/withdrawal |
| **Transaction data** | Transaction value, date/time, location, recipient, merchant, purpose | Welfare scheme documents specify the fields for each transaction record |
| **Behavioral profile** | Economic circumstances, behavioural patterns, consumption profile | Possible inference from transaction records |

**Table 3. Initial LLM Output for Data Attributes**

| Data category | Sensitivity | Volume | Origin | Purpose | Retention | Visibility | Intervenability |
|---|---|---|---|---|---|---|---|
| **Beneficiary identification data** | High | One record per beneficiary | Explicit disclosure by beneficiary/ identity infrastructure | Identify and verify the beneficiary | Not established | Scheme administrator, verification agencies, identity infrastructure; exact visibility unknown | Unknown |
| **Eligibility data** | High | One record per beneficiary | Government databases/ verification process | Determine eligibility and validate beneficiary information | Not established | Scheme administrator, eligibility-verification actors; exact visibility unknown | Unknown |
| **Payment data** | High | Many over payment lifecycle | Created /processed by CBDC /payment infrastructure | Transfer and administer welfare payment | Not established | CBDC/payment actors, potentially scheme administrator; exact visibility unknown | Unknown |
| **Transaction data** | High | Many transactions | Created by transaction processing; some attributes supplied by beneficiary /merchant | Execute and record transactions | Not established | Potentially multiple scheme/payment stakeholders; exact visibility unknown | Unknown |
| **Behavioral profile** | High | One evolving profile per beneficiary | Created by inference from transaction history | Potentially used for profiling or other purposes | Not established | Actors able to access the derived profile or underlying data; exact visibility unknown | Unknown |

## 5. Related Works

With the emergence of Large Language Models (LLMs), there has been an increasing focus on human-AI teaming, in which humans and AI are not separate actors but interdependent contributors who collaborate by complementing their strengths and compensating for their weaknesses (Baruwal Chhetri et al., 2024). LLMs can serve as flexible, on-demand cognitive aids that augment, rather than replace, human experts (Singh et al., 2025). Human-AI collaborative effort combining human intuition, contextual knowledge together with the computational capabilities and data processing power of AI may support in navigating complex situations (Albanese et al., 2025) and improve task performance (Tariq et al., 2025). Thus, rather than viewing AI as a replacement for human experts, AI can be envisioned as a teammate, and the value of collaboration depends on effective task allocation and integration of complementary capabilities (Kumar et al., 2026).

A Privacy Impact Assessment (PIA) is a process for examining the privacy implications of potentially privacy-invasive systems or services that process personal data (De & Le Metayer, 2016; Oetzel & Spiekermann, 2014). It is an inherently multidisciplinary activity requiring legal interpretation, technical analysis, domain-specific knowledge and stakeholder engagement. Consequently, PIAs present an ideal application area for human-AI teaming where AI can support knowledge-intensive tasks and human experts bring in contextual reasoning, regulatory judgment and accountability. Recent research has demonstrated that AI can effectively support individual privacy engineering tasks, including automated generation of Data Flow Diagrams (DFDs), preparation of structured threat models and attack trees and privacy risk categorisation (Zelenskiy & Sadovykh, 2025; Mollaeefar et al., 2025). However, these enhancements may lack the depth and reliability of traditional human expert driven privacy engineering activities (Zelenskiy & Sadovykh, 2025; Mollaeefar et al., 2025; Bissoli et al., 2026). The literature on human-AI collaboration for PIAs is in its nascent state. Privy, proposed by Lee et al. (2026) is a human-AI system that assists designers to envision and mitigate privacy risks in AI products and has led to further investigation on the broader role of human-AI teaming for privacy risk identification for AI-based products (Lee, 2026). These works are limited to envisioning privacy risks for AI products without any attempts to develop broader, comprehensive human-AI collaborative frameworks for PIAs. Many practical challenges that could emerge in human-AI collaborative PIAs also remain unresolved. These include identifying human-AI complementarity across PIA steps and orchestrating the collaborative workflow. Several PIA use cases involving potentially privacy-invasive applications and services have been discussed in the literature - fitness tracking services, online social networks, and contact tracing solutions (De & Le Metayer, 2022; De & Imine, 2020; Mollaeefar et al., 2025). Emerging domains, such as central bank digital currencies and their applications in welfare delivery, have received little attention from a PIA perspective, despite them potentially leading to novel privacy harms. These gaps call for further systematic research to develop a human-AI collaboration framework for PIAs, particularly in application domains that have received less attention in privacy risk assessment.

## 6. Conclusion and Future Work

In this paper, we propose a human–LLM teaming framework for the privacy risk analysis methodology PRIAM, specifying the task allocations for LLMs and human experts based on an integration of their complementary capabilities. The framework uses an LLM as an analytical resource, generating initial outputs for each activity through large-scale information extraction, synthesis, reasoning, and exploration. On the other hand, the framework uses human expertise to interpret and evaluate these outputs for their accuracy, assumptions, completeness, and supporting evidence, to direct refinements, and to retain final judgement about the output. We illustrate the framework on the data characterization activity of PRIAM for a CBDC-based welfare scheme use case. This illustration demonstrates how the LLM generates initial outputs on data categorization and assigns values for various attributes of each data category and how the human expert, based on this initial output, suggests refinements. This iterative approach of output generation and refinement can sharpen the output with appropriate evidence-based justification and identify unresolved gaps. The proposed framework can thus extend the depth and efficiency of the PRA, as LLMs synthesize large volumes of information sources more efficiently than humans, and human experts offer critical contextual evaluations and repeated feedback for refinements beyond LLM capabilities. While this work provides the foundation for human-LLM teaming for privacy risk analysis, future work can illustrate the proposed framework across all PRIAM activities and empirically examine whether and how the performance of human-LLM teams compares with human expert-only privacy risk analysis.

**References**

1. Albanese, M., Ou, X., Lybarger, K., Lende, D., Goldgof, D., Faisal, F. A., ... & Ghosh, A. (2025). Towards ai-driven human-machine co-teaming for adaptive and agile cyber security operation centers. *ACM Transactions on Internet Technology*.
2. Baruwal Chhetri, M., Tariq, S., Singh, R., Jalalvand, F., Paris, C., & Nepal, S. (2024). Towards human-AI teaming to mitigate alert fatigue in security operations centres. *ACM Transactions on Internet Technology*, *24*(3), 1-22.
3. Bissoli, A., Mollaeefar, M., Van Landuyt, D., & Ranise, S. (2026). Benchmarking the effectiveness of multi-agent LLMs in collaborative privacy threat modeling with LINDDUN GO. *Journal of Information Security and Applications*, *100*, 104489.
4. De, S. J. & Le Métayer, D. (2016). *Privacy Risk Analysis*. Morgan & Claypool Publishers.
5. De, S. J., & Imine, A. (2020). *Privacy risk analysis of online social networks*. Morgan & Claypool Publishers.
6. De, S. J. & Le Métayer, D. (2016). PRIAM: a Privacy Risk Analysis Methodology. In *Data Privacy Management and Security Assurance*. Springer.
7. Evertz, J., Chlosta, M., Schönherr, L., & Eisenhofer, T. (2024). Whispers in the machine: Confidentiality in llm-integrated systems. *arXiv preprint arXiv:2402.06922*.
8. Huang, Y., Song, J., Wang, Z., Zhao, S., Chen, H., Juefei-Xu, F., & Ma, L. (2025). Look before you leap: An exploratory study of uncertainty analysis for large language models. *IEEE Transactions on Software Engineering*, *51*(2), 413-429.
9. Kumar, M., Rani, R., Epiphaniou, G., & Maple, C. (2026). Adaptive Trust-Aware SOC Human–AI Teaming for resilient operations. *Computers & Security*, 104971.
10. Lee, H. P. H. (2026). Envisioning and Mitigating Privacy Risks for {Consumer-Facing}{AI} Product Concepts through {Human-AI} Teaming. In PEPR 2026: USENIX Conference on Privacy Engineering Practice and Respect.
11. Lee, H. P., Yang, Y. J., Bilik, M., Krsek, I., Von Davier, T. S., Monteiro, K., ... & Das, S. (2026, April). Privy: Envisioning and mitigating privacy risks for consumer-facing ai product concepts. In *Proceedings of the 2026 CHI Conference on Human Factors in Computing Systems* (pp. 1-30).
12. Mollaeefar, M., Bissoli, A., Van Landuyt, D., & Ranise, S. (2025, June). PILLAR: LINDDUN Privacy Threat Modeling Using LLMs. In *2025 IEEE European Symposium on Security and Privacy Workshops (EuroS&PW)* (pp. 278-286). IEEE.
13. Oetzel, M. C., & Spiekermann, S. (2014). A systematic methodology for privacy impact assessments: a design science approach. *European Journal of Information Systems*, *23*(2), 126-150.
14. Singh, R., Tariq, S., Jalalvand, F., Chhetri, M. B., Nepal, S., Paris, C., & Lochner, M.

(2026, May). Llms in the soc: An empirical study of human-ai collaboration in security operations centres. In *2026 IEEE Symposium on Security and Privacy (SP)* (pp. 4262-4281). IEEE.
15. Shi, X., Liu, J., Liu, Y., Cheng, Q., & Lu, W. (2025). Know where to go: Make LLM a relevant, responsible, and trustworthy searchers. *Decision Support Systems*, *188*, 114354.
16. Tariq, S., Chhetri, M. B., Nepal, S., & Paris, C. (2025). A2C: A modular multi-stage collaborative decision framework for human–AI teams. *Expert systems with applications*, *282*, 127318.
17. Wang, Y., Cai, C., Xiao, Z., & Lam, P. E. (2025). LLM access shield: domain-Specific LLM framework for privacy policy compliance. *arXiv preprint arXiv:2505.17145*.
18. Xia, Y., Kong, F., Yu, T., Guo, L., Rossi, R. A., Kim, S., & Li, S. (2024, May). Which llm to play? convergence-aware online model selection with time-increasing bandits. In *Proceedings of the ACM Web Conference 2024* (pp. 4059-4070).
19. Zelenskiy, L., & Sadovykh, A. (2025). Automating Cybersecurity Threat Analysis with LLMs. *ACM SIGAda Ada Letters*, *45*(2), 88-92.